\documentclass[
    ,final            
  ]
  {aipproc}

\layoutstyle{6x9}

\begin{document}

\title{Survival Probability at the LHC}

               

\author{Errol  Gotsman}{
  address={Department of Particle Physics, School of Physics and 
Astronomy, Raymond and Beverly Sackler Faculty of Exact Science, Tel Aviv
University, Tel Aviv 69978, Israel}
}
\begin{abstract}
Using a model based on two elements: the Good-Walker mechanism for 
low mass diffraction and multi-pomeron interactions for high mass 
diffraction, we obtain an excellent description of all aspects of soft 
scattering at high energy.   The parameters of the model 
 are determined by a fit to experimental data, giving the slope of 
the pomeron to be $\alpha'_{I\!\!P} \approx 0.01 \,GeV^{-2}$.
 We calculate the survival probability of  diffractive Higgs
production, and obtained a  value for this observable, which is smaller
than 1\% for the LHC energy range.
 \end{abstract}

\maketitle

\subsection{Introduction}
This talk is based on the contents of a recent paper by Gotsman, Levin, 
Maor and Miller \cite{GLMM}.

\par One of the key issues facing the high energy community is, whether 
the 
cornerstone of the Standard Model i.e. the Higgs boson will be discovered 
at the 
LHC. A suggested promising channel for its discovery is the diffractive 
process
$ p\;+\;p \rightarrow p\;+ H\;+\; p$, with rapidity gaps between the final 
protons and the Higgs boson.

\par We are interested in obtaining  
a reliable estimate of the probability of  seeing the Higgs at the LHC 
\cite{GLMM}.
The detection of the Higgs boson in the above channel depends crucially on 
the 
survival probability of the rapidity gap \cite{Bj93}, \cite{GLM1}, a 
calculation for which one requires knowledge of:
\begin{itemize}
\item the "hard" amplitude for Higgs production, which is a short distance 
process, and can be calculated using PQCD.
\item the survival factor of the gap also depends on  the "soft" elastic 
amplitude, 
which is believed to be a long distance process, and has to be evaluated 
using a model describing the relevant "soft" p-p interactions.
\end{itemize}

\par We determine the parameters of the soft amplitudes by making a fit to 
a data base containing all the relevant published data on p-p and 
$\bar{p}$-p interactions (see \cite{GLMM} for details). An unexpected 
result of our fit is that the value of the pomeron slope 
$\alpha'_{I\!\!P} \;=\;  0.012 \,GeV^{-2}$. This is consistent with the 
assumption of \cite{KMRNEW}, who take $\alpha'_{I\!\!P} \;=\; 0$.
 Our result suggests that the typical parton momentum is large
(approximately $<p_t> = 1/\sqrt{\alpha'_{I\!\!P} } \geq 7 \,GeV$). 
Therefore,
 the running QCD coupling $ \alpha'_{I\!\!P} \, =  \pi/b \ln( 
<p^2_t>/\Lambda^2_{QCD})  \,\ll \,1$
 (approximately  0.18), and we can consider it as
 our small parameter, when applying perturbative  QCD  estimates to
 the pomeron-pomeron interaction vertices.

\begin{figure}
  \includegraphics[height=.15\textheight]{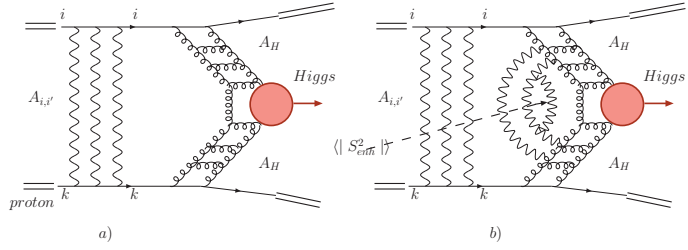}
  \caption{ Survival probability for exclusive central diffractive
production of the Higgs boson. (a) shows the contribution to
the survival probability in
the G-W mechanism, while (b)  illustrates the origin of the
additional factor
$\langle\mid S^2_{enh}\mid\rangle$. }
\end{figure}
 The details of the fit as well as the values of the parameters are given 
in \cite{GLMM}.

(1) In order to fit the single diffractive, double diffractive as well 
as the elastic amplitudes it is essential to include triple $ {I\!\!P}$
diagrams (for the large mass diffractive contribution) in addition to the 
usual 
Good-Walker (G-W) mechanism (which account only for the low mass 
diffractive contribution).

(2)  The fact that $\alpha'_{I\!\!P}$ is so small, encourages us to use 
PQCD as 
a guide for building a theory of pomeron interactions valid for all 
distances, and  there is no need for  separate "soft" and 
"hard" pomerons. 

(3) In the lowest order approximation of PQCD, only the transitions
${I\!\!P} \rightarrow 2 {I\!\!P}$ and  2${I\!\!P} \rightarrow {I\!\!P}$, 
should be considered, as  all other vertices are small. We, therefore 
restrict 
the summing of ${I\!\!P}$ diagrams to those with three ${I\!\!P}$ vertices 
only (i.e. fan diagrams).

(4) Since 4$\alpha'_{I\!\!P} ln(s/s_{0}) \ll$ 1 over the entire 
kinematic range, to simplify the algebra
 we take $\alpha'_{I\!\!P}$ = 0. 

(5) Details for summing the diagrams containing  multiple pomeron 
(enchanced) interactions, and obtaining  analytic expressions for the 
scattering amplitudes is given in \cite{GLMM}.
\vspace{-1.cm}
\subsection{Survival Probability of Diffractive Higgs Production }
In the following we  limit our discussion to the survival probability
of Higgs production, in an exclusive
central diffractive process.   Most estimates  of the values of survival
 probability have been made in the
framework of G-W mechanism, in two channel eikonal models.
A general review of such
survival probability calculations can be found in
\cite{heralhc}.
 The structure of the survival probability
expression is shown in Fig.1a, i.e.
\begin{equation} \label{SP}
\langle\mid S^2_{2ch} \mid \rangle = \frac{N(s)}{D(s)},
\end{equation}
where
\begin{eqnarray}
&N(s) = \int d^2\,b_1\,d^2\,b_2
\left[\sum_{i,k} \,<p|i>^2 <p|k>^2 \,A^{i}_H(s,b_1)\,A^k_H(s,b_2) 
(1-A^{i,k}_S ) \right]^2, \label{SP1}\\
&D(s) = \int\,d^2\,b_1\,d^2\,b_2
\left[\sum_{i,k} <p|i>^2 <p|k>^2\, A^i_H(s,b_1)\,A^k_H(s,b_2)\right]^2.
\label{SP2}
\end{eqnarray}
\vspace{-0.5cm}
 $<p|i>$ is  equal to $\langle \Psi_{proton}\mid \Psi_i \rangle$ hence , 
$<p|1> = \alpha$ and $ <p|2> = \beta$.
$A_S$ denotes the soft strong interaction  amplitude. 

The form of $ A_H(s,b)$ has been discussed in 
Refs.\cite{heralhc,SP2CH}.
In our model we assume an input Gaussian $b$ dependence
for the hard  amplitudes. We have
\begin{equation}\label{AH}
{A_{i,k}^H} = A_H(s)\,\Gamma_{i,k}^H(b),
\end{equation}
where $A_H(s)$ is an  $s$- dependent arbitrary function which does not
depend
on $i,k$, and \\
 $\Gamma_{i,k}^H(b) =
\frac{1}{\pi (R^H_{i,k})^2}\,e^{-\frac{2\,b^2}{(R^H_{i,k})^2}}$.
The hard vertices and radii ${R_{i,k}^H}^2$,
are constants derived from HERA $J/\Psi$ elastic and inelastic
photo and DIS production\cite{PSISL}.

Following Refs.\cite{heralhc,SP3P} we have introduced in the above,
two hard $b$-profiles
\begin{equation}
A^{pp}_H(b)\; = \;
\frac{V_{p \to p}}{2 \pi B_{el}^H}
 \exp \left( -\frac{b^2}{2\,B_{el}^H} \right) ; \;\;\;\;\;\; 
A^{pd}_H(b)\; = \; \frac{V_{p \to d}}{2 \pi
B_{in}^H} \exp \left( -\frac{b^2}{2 B_{in}^H}\right).
\label{2C11}
\end{equation}
 The values $B_{el}^H$=3.6 $GeV^{-2}$ and $B_{in}^H$=1 $GeV^{-2}$,
have been taken from the experimental ZEUS data on
$J/\Psi$ production at HERA (see Refs.\cite{heralhc,KOTE}).

\par
Eqn.(1) does not give a correct estimate for the survival
probability
and should be multiplied by a factor ($\langle\mid S^2_{enh} \mid 
\rangle$),  which
 incorporates the possibility for the Higgs boson to be emitted from
the enhanced diagrams (see fig1.b). Therefore, the resulting 
survival
 probability should be written as
\begin{eqnarray} \label{SPF}
\langle \mid S^2 \mid \rangle \,\,\,=\,\, \langle \mid S^2_{enh} \mid 
\rangle  
 \times\,\langle \mid S^2_{2ch} \mid \rangle.
\end{eqnarray}
  The results for the energy dependence of $\langle \mid S^2 \mid 
\rangle$, $\langle \mid S^2_{2ch} \mid \rangle$, and
$\langle \mid S^2_{enh} \mid \rangle$ are shown in Fig.2.

\begin{figure}
  \includegraphics[height=.3\textheight]{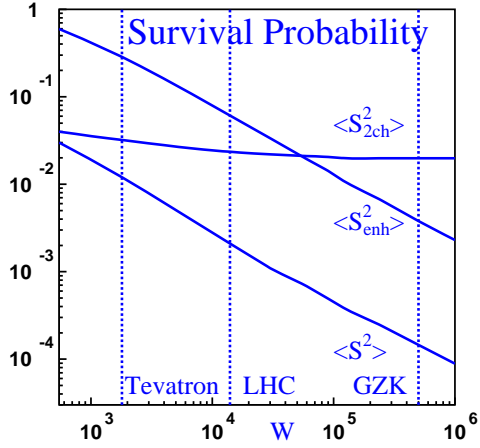}
  \caption{ Energy dependence of the survival probability for centrally 
produced Higgs. }
\end{figure}
\subsection{ Results, Discussion and Conclusions}
A paper by the Durham group (KMR) has recently been published 
\cite{KMRNEW},
 this paper has a similar philosophy to our paper \cite{GLMM}, 
however the 
conclusions of the two papers regarding  the values of the survival 
probability  at the LHC, differ greatly. 

The main differences are due to the fact that KMR:-

(1)  Do not carry out a fit to determine the values of their 
parameters, they prefer to "fine tune" these, and do not quote
a $\chi^{2}/d.f.$ for their model.

(2)  Use an adhoc assumption regarding the form of the multi-pomeron 
coupling i.e. they assume that $\Gamma[n  (I\!\!P) \rightarrow m (I\!\!P)] 
\; = \; mn \lambda^{m+n-2}$, where $\lambda$ is related to the difference 
between the absorptive cross section for an intermediate parton and that 
for the incoming proton. The above relation used  for multi-pomeron 
coupling, has its origin in the 
pomeron calculus of the 1970's, and its validity in the present context is 
questionable.

(3) Assume $S^{2}_{enh\;} \approx\;1$, (without any justification), and 
therefore their results for $S^{2}\;=\; S^{2}_{2 
channel}\times S^{2}_{enh}$, 
are much larger than ours.

\par Table 1 contains 
 a comparison of some of the results obtained in the two different 
approaches.

Bartels et al. \cite{Bar1} have also evaluated the gap survival 
probability (including the first pomeron loop) in the framework of pQCD, 
 our results for $S^{2}_{enh}\;$ and $S^{2}_{2ch}\;$ are very close to 
the 
values quoted in their paper, in spite of the very different formalism 
used. Consequently, Miller \cite{Mi}   evaluated the hard
  $\langle \mid S^2 \mid\rangle$, (including the first enhanced diagram) 
for the BFKL pomeron, and obtained a value of 0.4\%.   

We eagerly await results from the LHC, our calculations cast doubt that
the Higgs will be
discovered in the channel $ p + p \rightarrow p + H + p$.
\begin{table}
\begin{tabular}{||l|l|l|l||}
\hline 
  &  \,\,\,\,\,\,\,\,\,\,\,\,\,Tevatron & 
\,\,\,\,\,\,\,\,
\,\,\,\,\,\,\,\,LHC 
&\,\,\,\,\,\,\,\,\,\,\,\,\, $W= 10^5 GeV$  \\
 & GLM\,\,\,\,\,\,\,\,\,\,\,\,\,\,\,\,\,\,\,KMR & 
GLM\,\,\,\,\,\,\,\,
\,\,\,\,\,\,\,\,\,\,\,\,KMR & 
GLM\,\,\,\,\,\,\,\,\,\,\,\,\,\,\,\,\,
\,\,\,KMR\\\hline
$S^2_{2ch}(\%)$ & 3.2  \,\,\,\,\,\,\, 
\,\,\,\,\,\,\,\,\,\,\,\,2.7
 - 4.8 & 2.35  \,\,\,\,\,\,\,\,\,\,\,\, \,\,\,\,\,\,\,\,1.2-3.2 &
2.0  \,\,\,\,\,\,\,\,\,\,\,\, \,\,\,\,\,\,\,\,\,\,\,\, 0.9 - 2.5 \\
\hline
$S^2_{enh}(\%)$ & 28.5 \,\,\,\,\,\,\, 
\,\,\,\,\,\,\,\,\,\,100 & 6
.3  \,\,\,\,\,\,\,\,\,\,\,\, \,\,\,\,\,\,\,\,\,100 & 3.3 
\,\,\,\,\,\,\,
\,\,\, \,\,\,\,\,\,\,\,\,\,\,\,\,\,100 \\ \hline
$S^2(\%)$ & 0.91  \,\,\,\,\,\,\,\,\,\,\,\, \,\,\,\,\,2.7 
- 4.8
 & 0.15 \,\,\,\,\,\,\,\,\,\,\,\, \,\,\,\,\,\,\,\,1.2-3.2 &
0.066\, \,\,\,\,\,\,\,\,\,\,\,\, \,\,\,\,\,\,0.9 - 2.5\\
\hline 
\end{tabular}
  \caption{Comparison of results of the GLMM \cite{GLMM}  and KMR 
\cite{KMRNEW}
models.  }
\label{tab:a}
\end{table}
\subsection{Acknowledgements}
This research was supported
in part by the Israel Science Foundation, founded by the Israeli Academy 
of Science and Humanities,
 by BSF grant $\#$ 20004019 and by
a grant from Israel Ministry of Science, Culture and Sport and
the Foundation for Basic Research of the Russian Federation.


\vspace{-0.5cm}
\bibliographystyle{aipproc}   

\bibliography{sample}

\begin{thebibliography}{9}
\bibitem{GLMM}
E.~Gotsman, E.~Levin, U.~Maor and J.~S.~Miller,  \emph{Eur.Phys.J.C}, 
(in print),arXiv:0805.2799[hep-ph]. 
 
\bibitem{Bj93}
J.~D.~Bjorken, \emph{Int. J. Mod. Phys.} \textbf{A7},4189 (1992);  
\emph{Phys.Rev.} \textbf{D47},101 (1993).

\bibitem{GLM1}
E.~Gotsman, E.~Levin and U.~Maor, \emph{Phys.Rev.} \textbf{D49},R4321 
(1994).
 
\bibitem{KMRNEW}
M.~G.~Ryskin, A.~D.~Martin and V.~A.~Khoze, \emph{Eur.Phys.J.}
\textbf{C54}, 199 (2008).




\bibitem{heralhc}
E.~Gotsman, E.~Levin, U.~Maor, E.~Naftali, A.~Prygarin, in \emph{HERA and 
the LHC:  A workshop on the Implications of HERA for LHC Physics:} 
\textbf{Proceedings Part A},221(2005).

\bibitem{SP2CH}
E.~Gotsman, E.~Levin and U.~Maor, \emph{Phys.Rev.} \textbf{D60}, 094011
(1999).

\bibitem{PSISL}
Zeus Collaboration: \emph{Nucl.Phys.} \textbf{B695}, 3 (2004): 
\emph{Eur.Phys.J.} \textbf{C24}, 345 (2004).

\bibitem{SP3P}
E.~Gotsman, A.~Kormilitzin, E.~Levin and U.~Maor, \emph{Eur.Phys.J.}, 
\textbf{C52},295 (2007).

\bibitem{KOTE}
H.~Kowalski, D.~Teaney, \emph{Phys.Rev.} \textbf{D68},114005 (2003). 
  
\bibitem{Bar1}
J.~Bartels, S.~Bondarenko, K.~Kutak and L.~Motyka, \emph{Phys.Rev.} 
\textbf{D73}, 093004 (2006).

\bibitem{Mi}
J.S.~Miller, \emph{Eur.Phys.J.}, \textbf{C56},39 (2008).



\end{thebibliography}


\end{document}